\documentclass[aps,twocolumn,superscriptaddress]{revtex4}

\usepackage{graphicx}

\begin{document}
\bibliographystyle{apsrev}


\title{The electronic state of vortices in YBa$_2$Cu$_3$O$_y$ investigated by complex surface impedance measurement}

\author{Yoshishige Tsuchiya}
\email{tsuchi@maeda1.c.u-tokyo.ac.jp}
\author{Katsuya Iwaya}
\author{Kentarou Kinoshita}
\affiliation{Department of Basic Science, University of Tokyo,\\
3-8-1, Komaba, Meguro-ku, Tokyo 153-8902, Japan}

\author{Tetsuo Hanaguri}
\affiliation{Department of Advanced Materials Science, University of Tokyo,\\
7-3-1, Hongo, Bunkyo-ku, Tokyo 113-8656, Japan}
\affiliation{CREST, Japan Science and Technology Corporation (JST), \\ 4-1-8, Honcho, Kawaguchi, 332-0012, Japan}

\author{Haruhisa Kitano}
\author{Atsutaka Maeda}
\email{maeda@maildbs.c.u-tokyo.ac.jp}
\affiliation{Department of Basic Science, University of Tokyo,\\
3-8-1, Komaba, Meguro-ku, Tokyo 153-8902, Japan}
\affiliation{CREST, Japan Science and Technology Corporation (JST), \\ 4-1-8, Honcho, Kawaguchi, 332-0012, Japan}

\author{Kenji Shibata}
\affiliation{Institute for Materials Research, Tohoku University,\\
2-1-1 Katahira, Aoba-ku, Sendai 980-8577, Japan}

\author{Terukazu Nishizaki}
\author{Norio Kobayashi}
\affiliation{Institute for Materials Research, Tohoku University,\\
2-1-1 Katahira, Aoba-ku, Sendai 980-8577, Japan}
\affiliation{CREST, Japan Science and Technology Corporation (JST), \\ 4-1-8, Honcho, Kawaguchi, 332-0012, Japan}

\date{\today}

\begin{abstract}
The electromagnetic response to microwaves in the mixed state of YBa$_2$Cu$_3$O$_y$(YBCO) was measured in order to 
investigate the electronic state inside and outside the vortex core. 
The magnetic-field dependence of the complex surface impedance at low temperatures was in good agreement with a general 
vortex dynamics description assuming that the field-independent viscous damping force and the linear restoring force were 
acting on the vortices. 
In other words, both real and imaginary parts of the complex resistivity, $\rho_1$, and $\rho_2$, were linear in $B$. 
This is explained by theories for $d$-wave superconductors. 
Using analysis based on the Coffey-Clem description of the complex penetration depth, we estimated that the vortex 
viscosity $\eta$ at 10 K was (4 $\sim$ 5) $\times$ 10$^{-7}$ Ns/m$^2$. 
This value corresponds to $\omega_0 \tau \sim$ 0.3 - 0.5, where $\omega_0$ and $\tau$ are the minimal gap frequency and 
the quasiparticle lifetime in the vortex core, respectively. 
These results suggest that the vortex core in YBCO is not in the deeply superclean regime but in the moderately clean 
regime. 
Investigation of the moderately clean vortex core in high-temperature superconductors is significant because physically 
new effects may be expected due to $d$-wave characteristics and to the quantum nature of cuprate superconductors. 
The behavior of $Z_s$ as a function of $B$ across the first order transition (FOT) of the vortex lattice was also 
investigated. 
Unlike Bi$_2$Sr$_2$CaCu$_2$O$_y$ (BSCCO), no distinct anomaly was observed around the FOT in YBCO. 
Our results suggest that the rapid increase of $X_s$ due to the change of superfluid density at the FOT would be observed 
only in highly anisotropic two-dimensional vortex systems like BSCCO. 
We discuss these results in terms of the difference of the interlayer coupling and the energy scale between the two 
materials. 
\end{abstract}

\maketitle

\section{INTRODUCTION}

The properties of high-$T_c$ superconductors (HTSC) in an applied magnetic field have revealed a rich variety of 
fascinating phenomena. 
In particular, investigations of the electronic structure in the mixed state of HTSC have been key experiments in recent 
condensed matter physics. 
In conventional superconductors (CSC) with $s$-wave gap symmetry, there are bound states of quasiparticles (QPs) localized 
in the vortex core. 
The level separation between the bound states, $\Delta E$, is
\begin{equation}
\Delta E \equiv \hbar \omega_0 \sim \Delta_0/k_F \xi, 
\end{equation}
where $\Delta_0$ is the bulk gap energy and $k_F$ and $\xi$ are the Fermi wavenumber and the GL coherence length, 
respectively~\cite{CdGM}. 
Here, $\omega_0$ is regarded as the minimal gap frequency inside the core. 

In CSC, $T_c$ is below 10 K and $\xi \sim$ 100 \AA, which lead to $\Delta E \le$ 0.1 K. 
This could not be resolved by available experimental techniques. 
Indeed, in the scanning tunneling spectroscopy (STM) experiments in CSC~\cite{Hess89,Renner91}, the local density of 
states inside the core was shown as a zero energy peak anomaly, which was interpreted as a smearing of the discrete levels 
in the core~\cite{ShorePL,GygiPBR90}. 

When describing the dynamics of QPs in the mixed state, the key parameter is the width of the level in the core, $\delta 
E$, which is connected with the scattering time, $\tau$, of the QPs in the vortex core. 
The relationship between $\delta E$ and $\tau$ was expressed as
\begin{equation}
\delta E \sim \frac{\hbar}{\tau}. 
\end{equation}
Since $\delta E \gg \Delta E$ in CSC, the vortices in CSC can be regarded as tubes with a normal metal of radius $\xi$. 
In such a case, the vortex motion was well described with the Bardeen - Stephen (B-S) model~\cite{BarSte}. 

In HTSC, such a simple description of the vortex core is incorrect for several reasons. 
First, the energy gap has $d$-wave symmetry in HTSC~\cite{HarlingenRMP}. 
Since the amplitude of the gap at the node is zero, QPs are not localized in the vortex core but extended along the node 
direction~\cite{Volo93,SchoMaki,Ichioka}. A theoretical calculation suggested that there were no truly localized states in 
the vortex core in pure $d_{x^2-y^2}$ superconductors~\cite{WanMac}. 
On the other hand, recent STM results in HTSC showed that the peak at zero energy vanished and that the density of states 
at a finite energy below the gap $\Delta_0$ was enhanced near the center of the vortex core~\cite{Maggio, Renner, PanPRL}. 
Therefore, the presence or the absence of localized states in the vortex core of HTSC is one of the central basic 
questions. 

The second different feature of the vortex core in HTSC is the semi-quantum nature of the core. Since $\Delta_0 \sim$ 20 
meV for HTSC, that means $\xi \sim$ 20 \AA. This gives $k_F \xi \sim 20$, which is quite different from the case of CSC 
(for which $k_F \xi \sim$ 200, typically). 
Indeed, if the peak observed in the STM experiment mentioned above corresponds to the quantum levels in the core, it gives 
$k_F \xi \sim 2$. 
The physics of such a quantum core has been undeveloped up to now. 

Such anomalous aspects of the electronic structure around the core in HTSC should affect the dynamic response of the 
vortices. 
If the vortex motion is described phenomenologically, the dissipation can be discussed in terms of the frictional force, 
\begin{equation}
\mathbf{F}_v = \eta \mathbf{v}_L + \alpha_H \mathbf{v}_L \times \hat{\mathbf{z}}, 
\end{equation}
which is proportional to the velocity of vortices, $\mathbf{v}_L$. 
Here $\eta$ and $\alpha_H$ are the viscous drag and the Hall coefficients, respectively, and $\hat{\mathbf{z}}$ is the 
unit vector parallel to an applied magnetic field. 
The dissipation mechanism is, in turn, closely related to the characteristic scattering time of QPs in the vortex core, 
$\tau$. 
The vortex viscosity $\eta$, and Hall constant $\alpha_H$ were calculated~\cite{KopKraJL76,BlatterRMP} for arbitrary 
values of $\omega_0 \tau$, and it was found that
\begin{eqnarray}
\eta &=& \pi \hbar n\frac{\omega_0 \tau}{1+(\omega_0 \tau)^2}\label{eq:Blaeta}, \\  \alpha_H &=& \pi \hbar 
n\frac{(\omega_0 \tau)^2}{1+(\omega_0 \tau)^2}\label{eq:Blaalpha}, 
\end{eqnarray}
where $n$ is the QP concentration. 
The ratio of the level spacing $\Delta E$ to the level width $\delta E$, 
\begin{equation}
\frac{\Delta E}{\delta E} = \omega_0 \tau, 
\label{eq:ratio}
\end{equation}
is used as an index of the cleanness in the vortex core. 
In the dirty limit ($\omega_0 \tau \ll 1$), the Hall effect (Eq.~(\ref{eq:Blaalpha})) is negligible, and 
Eq.~(\ref{eq:Blaeta}) was expressed as $\eta = \pi \hbar n \omega_0 \tau = B_{c2} \Phi_0/\rho_n$, where $\rho_n = 
m^*/ne^2\tau$, and $B_{c2}$ is the upper critical field. 
This is the B-S expression of the viscosity, $\eta$~\cite{BarSte}. 
In the opposite limit, $\omega_0 \tau \gg 1$(the superclean limit), the Hall effect is dominant and $\eta$ is expressed as 
$\eta \simeq \pi \hbar n/\omega_0 \tau$. 
According to~\cite{GoloReview}, the diagonal component of the vortex resistivity under constant transport current, 
$\mathbf{J}$, was expressed as
\begin{equation}
\rho_v = \frac{B\phi_0}{\eta + \alpha_H^2/\eta} = \frac{B \phi_0}{\eta_{\rm eff}}, 
\label{eq:rhoxx}
\end{equation}
where the effective viscosity $\eta_{\rm eff}$ was defined as 
\begin{equation}
\eta_{\rm eff} = \eta + \alpha_H^2/\eta. 
\label{eq:etaeff}
\end{equation}
This means that the experimentally obtained viscous coefficient under the condition $\mathbf{J} = {\rm const.}$, was 
$\eta_{\rm eff}$. 
Equations~(\ref{eq:Blaeta}), (\ref{eq:Blaalpha}), and (\ref{eq:etaeff}) yield
\begin{equation}
\omega_0 \tau = \frac{\eta_{\rm eff}}{\pi \hbar n}. 
\label{eq:etatau}
\end{equation}
Therefore, we can determine the QP lifetime in the vortex core from estimates of the viscosity of the vortex motion in the 
mixed state of HTSC. 

Due to the strong pinning in HTSC, dc-measurements are not appropriate to study the vortex dynamics at low temperatures. 
Instead, measurement of the microwave electromagnetic response is a powerful technique to probe the dynamics of the 
vortices in HTSC. 
From the high-frequency response of the vortices, several parameters relating to the vortex dynamics, such as the 
viscosity, pinning constant, and Hall constant can be estimated according to existing theories of vortex motion. 

There have already been many experimental efforts to determine the vortex dynamics parameters of YBa$_2$Cu$_3$O$_y$, and a 
thorough review is provided in~\cite{GoloReview}. 
To the best of our knowledge, experimental determination of the vortex viscosity $\eta$ in YBCO at low temperatures was 
first performed by Pambianchi {\it et al.}, using the parallel plate resonator technique~\cite{PambiIEEE93}. 
They measured the magnetic-field dependence of the complex resistivity up to 0.4 T at 11 GHz and showed that the 
temperature dependence of $\eta$ was consistent with a temperature-dependent normal state resistivity in the context of 
the B-S model~\cite{BarSte}. 
A similar result was obtained in~\cite{Morgan94} by measuring the surface resistance $R_s$ as a function of magnetic field 
up to 8 T at several frequencies. 
A different report on the $R_s$ measurement as a function of magnetic field up to 7 T at low 
temperatures~\cite{Matsuda94}, deduced a very large viscosity, which was two orders of magnitude larger than in 
conventional superconductors. 
This huge value was interpreted as evidence that the core of YBCO was in the superclean regime, $\omega_0 \tau \gg 1$. 
However, their estimation of the viscosity was based on the assumption that the pinning frequency $\omega_p$ is much 
smaller than the measurement frequency $\omega$. 
Under that assumption, they estimated the viscosity only from the data of the surface resistance, $R_s$ (the real part of 
the surface impedance, $Z_s$). 
In CSC, the assumption, $\omega \gg \omega_p$ was reasonable in the microwave region since the characteristic pinning 
frequency $\omega_p \sim$ 100 MHz~\cite{GittRosen}. 
In HTSC, however, the condition $\omega \gg \omega_p$ might not be satisfied. 
Indeed, experimentally estimated values of the pinning frequency $\omega_p$ at low temperatures in previous 
reports~\cite{PambiIEEE93,Revenaz94,GoloPRB94} were on the order of $\sim$ 10 GHz. 
In such a case, the estimation of $\eta$ from $R_s$ alone leads to an incorrect value for $\eta$, and the measurement of 
$Z_s$ as a complex value is essential. 
Parks {\it et al.}~\cite{Parks95} measured the complex resistivity, $\tilde{\rho}$, of YBCO thin films as a function of 
frequency in the THz domain. 
 They reported a relatively small $\eta$ and insisted that the smallness was due to the effects of anisotropy and $d$-wave 
gap symmetry. 
 If their interpretation is correct, there should be a definite difference in the magnetic-field dependence of the surface 
impedance from what would be expected from the usual flux-flow theory. 
However, they reported only the frequency dependence of the complex resistivity $\tilde{\rho}(\omega)$ at a fixed field 6 
T, and detailed field dependence data were unavailable. To discuss the deviation from the conventional vortex dynamics 
description, the magnetic-field dependence of $\tilde{\rho}$ should be investigated. 

Considering the state of previous high-frequency measurements of $\eta$-values, there is a large ambiguity. 
To resolve such confusion, complex surface impedance measurements as functions of magnetic field, frequency, and 
temperature should be performed up to higher magnetic fields. 

In addition, it should be emphasized again that in HTSC, the vortex core is likely to be close to the quantum limit since 
$k_F \xi$ is small and the major component of the condensate wave function has $d$-wave symmetry. 
In such a highly anomalous situation, to the best of our knowledge, there have been no rigorous calculations of the 
dynamic properties of vortices. 
So the appearance of some new effect may be expected. 

Another problem we focus on in this study is the relationship between the electronic structure and the vortex structure in 
the mixed state. 
Previously, we found that in Bi$_2$Sr$_2$CaCu$_2$O$_y$, a rapid increase of surface reactance, $X_s$, took place at the 
first order transition (FOT) of the vortex lattice~\cite{HanaguriPRL}. 
Since $X_s$ is proportional to the real part of the effective penetration depth, it was proposed that this increase may be 
ascribed to the decrease in the superfluid density rather than to the loss of pinning. 
 This suggested that the change in the structure of the vortex lattice largely affected the electronic structure of the 
vortex. 
If this phenomenon were observed in other cuprates, where the FOT occurred, it would be a universal new effect found in 
HTSC associated with the FOT of the vortex lattice.
Thus, we have investigated how universal the phenomenon is in materials with different values of anisotropy. 

Furthermore, quite recently, the existence of a collective mode in the microwave frequency range was predicted in 
unconventional superconductors with mixed symmetry order parameters~\cite{Balatsky00}. 
Such a mode might be expected to show up as a resonance in the surface impedance. 

To study the fascinating problems described above, in this paper, we present {\it complex} surface impedance measurements 
on untwinned YBCO single crystal as functions of temperature and magnetic field. 
These measurements have been made in the wide region of the mixed state (10 K $< T < T_c$, 0 $< H <$ 15 T) at several 
frequencies, and provide information on electronic structure in the mixed state. 
In the following section, details of the experimental technique to measure $Z_s$ in an applied magnetic field together 
with the analysis based on vortex dynamics will be described. 
 In Sec. III, the results of the $Z_s$ measurements will be shown. 
 Both the real and imaginary parts of the complex resistivity show the $B$-linear behavior at low temperatures below 60 K. 
 Through comparison with a theoretical calculation by Coffey and Clem~\cite{CC}, we estimate the pinning frequency, vortex 
viscosity and pinning constant. 
Estimated values of $\eta$ at 10 K are $\sim$ 4 - 5 $\times$ 10$^{-7}$ Ns/m$^2$. These values correspond to $\omega_0 \tau 
\sim$ 0.3 - 0.5, which suggests that the vortex core of YBCO is not in the deeply superclean regime but in the moderately 
clean regime. 
 In the data from the high temperature region above 60 K, we cannot find any obvious anomaly around the FOT in both $R_s$ 
and $X_s$. 
  This is in contrast to the results observed in BSCCO. 
 In Sec. IV, the observed $B$-linear behavior is explained in terms of a theoretical consideration for $d$-wave 
superconductors. 
 The significance of the moderately clean vortex core realized in YBCO will be discussed together with the relationship 
between the $Z_s$ and the FOT in YBCO, and the difference between YBCO and BSCCO. 
Moreover, we will briefly comment on the possibility of the existence of the collective mode predicted by Balatsky {\it et 
al.}. 
Sec. V is dedicated to our conclusions. 

\section{EXPERIMENTAL}

\subsection{Apparatus}

YBa$_2$Cu$_3$O$_y$ single crystals were grown by a self-flux method using Y$_2$O$_3$ crucibles and were annealed at 
450$^{\circ}$C. 
The sample described in this study was a naturally untwinned, slightly overdoped single crystal with dimensions of 0.6 
$\times$ 0.6 $\times$ 0.05 mm$^3$. 
It showed a superconducting transition at $T_c$ = 91.2 K, which was determined from the magnetization measurement using a 
commercial SQUID magnetometer (Quantum Design). 

The microwave surface impedance $Z_s \equiv R_s +iX_s$, where $R_s$ and $X_s$ are the surface resistance and surface 
reactance, respectively, was measured by the cavity perturbation technique with cylindrical cavities operated in the 
TE$_{011}$ mode. 
Since a superconducting cavity with high sensitivity~\cite{SriKenRSI88} could not be used in an applied magnetic field, we 
used copper cavities tuned to 19.1, 31.7, and 40.8 GHz. 
The $Q$-values of each cavity at 10 K were $\sim 6.5 \times 10^4$ at 19.1 GHz, $\sim 2.8 \times 10^4$ at 31.7 GHz, and 
$\sim 2.7 \times 10^4$ at 40.8 GHz. 
The sample was located at the center of each cavity, at the antinode of the microwave magnetic field $H_{\rm rf}$ being 
parallel to the $c$-axis of the sample. 
Thus, the microwave currents flowed in the $ab$ plane (CuO$_2$ plane). 
Details will be described in a separate publication. 
$R_s$ and $X_s$ for the sample in zero field could be simultaneously obtained from the shift in the quality factor of the 
cavity, $\Delta (1/2Q)$, and the shift in the resonance frequency, $\Delta f$. 
According to cavity perturbation theory~\cite{KleinIJIMW1}, in the skin depth regime where the microwave skin depth is 
significantly shorter than the dimensions of the sample, $R_s$ and $X_s$ may be expressed as 
\begin{eqnarray}
R_s &=& G\left(\frac{1}{2Q_s} - \frac{1}{2Q_b}\right) \equiv G\Delta \left(\frac{1}{2Q}\right)\label{eq:rsq} \\ X_s &=& G\left(-\frac{f_s - f_b}{f_b}\right)
+C \equiv G\left(-\frac{\Delta f}{f_b}\right)+C\label{eq:xsf},
\end{eqnarray}
where $Q_s$ and $f_s$ are the quality factor and the resonant frequency of the cavity including the sample, and $Q_b$ and 
$f_b$ are those of the empty cavity without the sample. $G$ is a geometrical factor that depends only on the shape of the 
sample and the cavity, and $C$ is a metallic shift constant, which represents the shift of the resonance frequency of the 
cavity with a perfect conductor with the same dimensions as the sample. 
In zero field, $X_s$ is proportional to the London penetration depth, $\lambda_L$, as $X_s = \mu \omega \lambda_L$, where 
$\mu$ is the vacuum permeability and $\omega$ is the angular frequency. 

However, in Eq. (\ref{eq:xsf}), the effects of thermal expansion should be taken into account at high temperatures, as was 
pointed out in~\cite{DresselIJIMW3}. 
Therefore, Eq.~(\ref{eq:xsf}) were corrected to the following expressions,
\begin{equation}
X_s = G\left(\left(-\frac{\Delta f}{f_b}\right)\alpha(T)\right)+C\label{eq:xsfc}, 
\end{equation}
where $\alpha (T)$ is the temperature-dependent factor related to the relative difference of thermal expansivities between 
the cavity and the sample. 
In order to determine the absolute values of $R_s$ and $X_s$, we should fix the two parameters $G$, and $C$, and estimate 
$\alpha(T)$ in Eqs.~(\ref{eq:rsq}) and~(\ref{eq:xsfc}). 
We determined $G$, $C$, and $\alpha (T)$, as follows. 
First, in the normal state, the relationship, $R_s = (\mu\omega\rho_n/2)^{1/2}$ is appropriate. 
$\rho_n$ of the sample at 100 K is 6.4 $\times$ 10$^{-7}$~$\Omega$ m, which gives $R_s = 0.28$~$\Omega$ at 31.7 GHz. 
Substitution of the value into Eq.~(\ref{eq:rsq}) gives the value of $G$. 
Second, $\alpha(T)$ was estimated from the difference of the thermal expansivities between Cu~\cite{KroegerJAP77}, and 
YBCO~\cite{KrautPC93}. 
Finally, the value of $C$ was determined by assuming that the London penetration depth at $T =$ 0 K, $\lambda_L(0) \sim$ 
1400 \AA ~\cite{AnlagePBR91}. 
The corrected $X_s$ data satisfied that (1) the Hagen-Rubens relation ($R_s = X_s$) was held in the normal state above 
$T_c$, and (2) at $T \to$ 0, $X_s/\mu \omega$ was extrapolated to $\lambda_L(0)$. 
After this correction, the temperature dependence of $\lambda_L$ in zero field was consistent with the data reported 
in~\cite{HossePRB99}. 
$R_s$ and $X_s$ in an applied magnetic field were subsequently determined by using the values of $G$, $C$, and $\alpha(T)$ 
in zero field. 
We confirmed that the magnetic field dependence of $Z_s$ was not affected by the corrections mentioned above. 
Surface impedance measurements were performed both with swept $H$ after zero-field cooled conditions and with swept $T$ 
under field-cooled conditions. 
In all measurements, the static applied magnetic field $H_{\rm dc}$ was parallel to the $c$-axis. 

\subsection{Method of the Analysis}

In this subsection, we will explain the procedure used in this study to determine the vortex dynamics parameters. 
The complex surface impedance $Z_s$ is related to the complex penetration depth, $\tilde{\lambda}$, as 
\begin{equation}
Z_s=i\mu \omega \tilde{\lambda}, 
\end{equation}
The complex resistivity $\tilde{\rho} = \rho_1 + i\rho_2$ is expressed as 
\begin{equation}
\tilde{\rho} = Z_s^2/i\mu \omega = i\mu \omega \tilde{\lambda}^2. 
\label{eq:crho}
\end{equation}
The general behavior of $\tilde{\lambda}$ in the mixed state of type-II superconductors was calculated by Coffey and Clem 
(C-C)~\cite{CC}. 
In this model, a rigid, massless vortex lattice with linear restoring force and viscous damping force was assumed. 
The basic equation of motion of vortex was described as 
\begin{equation}
\eta \dot{\mathbf{u}} + \kappa \mathbf{u} = \phi_0 \mathbf{J} \times \hat{\mathbf{z}} + \mathbf{f}(t) 
\label{eq:eom}
\end{equation}
where $\mathbf{u}$ is the displacement of vortices, $\eta$ is the viscosity in the absence of flux creep, $\kappa$ is the 
pinning constant, $\mathbf{J}$ is the transport current density, $\phi_0$ is flux quantum and $\hat{\mathbf{z}}$ is the 
unit vector parallel to $H_{\rm dc}$. 
$\mathbf{f}(t)$ is the stochastic force, which corresponds to the effects of creep, and the Hall effect was neglected. 
Based on a self-consistent treatment of vortex dynamics including the influence of the normal component in the two-fluid 
sense, they derived an expression for the complex penetration depth $\tilde{\lambda}$ as a function of temperature, 
magnetic-field, and frequency, 
\begin{eqnarray}
\frac{\tilde{\lambda}^2}{\lambda_L^2} &=& \left( \frac{1}{1+s^2}+\frac{(1-\epsilon)r-s(\epsilon+r^2)}{(1+s^2)(1+r^2)}b 
\right)\nonumber \\  &-& i \left( \frac{s}{1+s^2}+\frac{(\epsilon+r^2)+s(1-\epsilon)r}{(1+s^2)(1+r^2)}b \right) \label{eq:lamall}. 
\end{eqnarray}
Here, $s = 2\lambda_L^2/\delta_{nf}^2$, where $\lambda_L$ is the London penetration depth and $\delta_{nf}$ is the normal 
fluid skin depth ($\delta_{nf}^2 = 2\rho_n/\mu \omega$). 
The normalized field $b$, and normalized frequency $r$ were expressed as, 
\begin{equation}
b=\phi_0B/\mu \omega \lambda_L^2\eta = B/B_\eta, 
\label{eq:beta}
\end{equation}
\begin{equation}
r=\omega/\omega_p, 
\label{eq:r}
\end{equation}
where $\omega_p=(\kappa_p/\eta)(I_1(\nu)I_0(\nu))/(I_0(\nu)-1), \nu=U/k_BT$, $U$ is an activation energy of the pinning 
potential, $\epsilon =1/I_0(\nu)$, and $I_n$ is a modified Bessel function of the first kind of the order $n$. 

At low temperatures, ($T \le$ 60 K), which corresponds to $s \to 0$ and $\nu \to \infty$, i.e. $\epsilon \to 0$, effects 
of both the flux creep and the thermally activated normal fluid can be neglected, and the equation~(\ref{eq:lamall}) 
becomes a rather simple expression, 
\begin{equation}
\frac{\tilde{\lambda}^2}{\lambda_L^2} = \left( 1+\frac{r}{1+r^2}b \right) -i \left( \frac{r^2}{1+r^2}b \right). 
\label{eq:lam}
\end{equation}
$R_s$ and $X_s$ normalized by $\mu \omega \lambda_L$ are deduced from Eq.~(\ref{eq:lam}), and expressed as
\begin{eqnarray}
\frac{R_s}{\mu \omega \lambda_L} &=& \frac{1}{2} \left( \sqrt{1+\frac{2r}{1+r^2}b+\frac{r^2}{1+r^2}b^2}+\frac{r^2}{1+r^2}b 
\right)^{\frac{1}{2}} \nonumber \\ &-& \frac{1}{2} \left( 
\sqrt{1+\frac{2r}{1+r^2}b+\frac{r^2}{1+r^2}b^2}-\frac{r^2}{1+r^2}b \right)^{\frac{1}{2}},\label{eq:rs} \\
\frac{X_s}{\mu \omega \lambda_L} &=& \frac{1}{2} \left( \sqrt{1+\frac{2r}{1+r^2}b+\frac{r^2}{1+r^2}b^2}+\frac{r^2}{1+r^2}b 
\right)^{\frac{1}{2}} \nonumber \\ &+& \frac{1}{2} \left( 
\sqrt{1+\frac{2r}{1+r^2}b+\frac{r^2}{1+r^2}b^2}-\frac{r^2}{1+r^2}b \right)^{\frac{1}{2}}\label{eq:xs}. 
\end{eqnarray}
Note that Eqs. (\ref{eq:lam}), (\ref{eq:rs}), (\ref{eq:xs}) contain only two parameters, $r$ and $b$. 
Variability of parameters $r$ and $b$ is limited within the range where C-C model is valid ($B_{c1} \ll B \ll B_{c2}$, where $B_{c1}$ and $B_{c2}$ are lower and upper critical field, respectively, and $\hbar \omega < \Delta_0$). 
Therefore, ranges of $r$ and $b$ are $0 \ll r \ll \infty$, and $B_{c1}/B_{c2}(\sim 0) \ll b \ll B_{c2}/B_{c1}(\sim \infty)$, respectively. 
\begin{figure}[bp]
\begin{center}
\includegraphics[height=15pc]{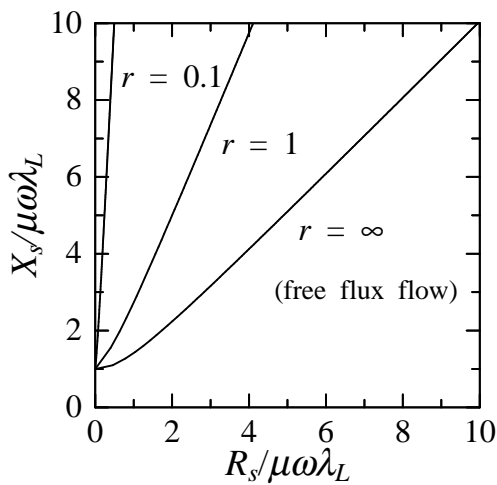}
\end{center}
\caption{Behavior of Eqs.~(\ref{eq:rs}) and (\ref{eq:xs}) in ``the impedance plane plot" at several fixed $r$ values. The 
curve, $r = \infty$, represents the free flux flow state. }
\label{fig:IP}
\end{figure}

Here we introduce ``the impedance-plane plot", where the horizontal axis is $R_s/\mu \omega \lambda_L$ and the vertical 
axis is $X_s/\mu \omega \lambda_L$, as is shown in Fig.~\ref{fig:IP}. $R_s/\mu \omega \lambda_L$ and $X_s/\mu \omega 
\lambda_L$ of Eq.~(\ref{eq:rs}), and~(\ref{eq:xs}) were plotted through the parameter $b$ with fixed $r$'s. 
In the figure any curves in this plane show a concave upward curvature. 
Therefore, provided that the experimental data obeys the behavior in Fig~\ref{fig:IP}, we can estimate the depinning 
frequency $\omega_p$ uniquely by comparing the measured $Z_s (B)$ data with the curves in Fig.~\ref{fig:IP}. 
After that we can determine $\eta$ through Eqs.~(\ref{eq:rs}), and~(\ref{eq:xs}). 
To sum up the procedure, first, we try to fit the data with the curve in Fig.~\ref{fig:IP}, for a fixed $r$ value.
Substituting the $r$ value into Eqs. (\ref{eq:rs}) and (\ref{eq:xs}) enables us to determine another parameter $b$.
Finally, the comparison between $b$ and experimental $B$ gives the $\eta$ value.\begin{figure}[htbp]
\begin{center}
\includegraphics[height=25pc]{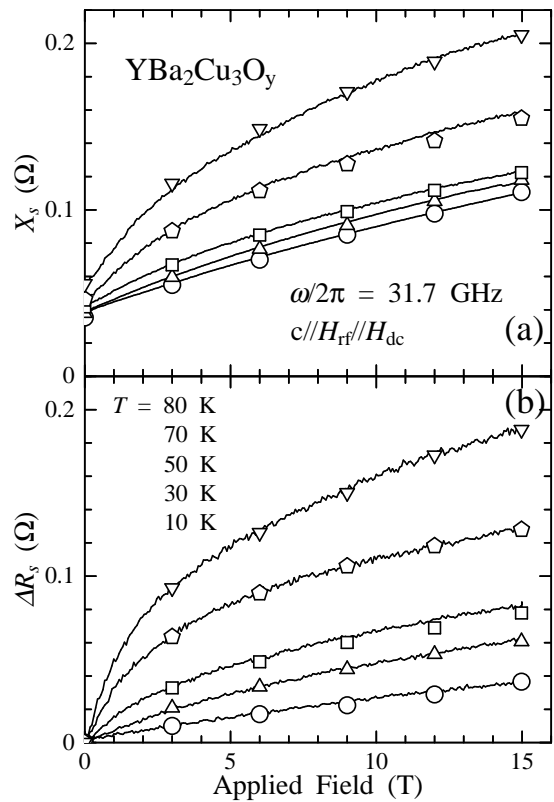}
\end{center}
\caption{Magnetic field dependence of (a) $X_s$ and (b) $\Delta R_s$ of YBCO up to 15 T at 31.7 GHz at various 
temperatures. Continuous lines and open marks represent the results from swept magnetic-field measurements and swept 
temperature measurements, respectively. }
\label{fig:RXTH}
\end{figure}

\section{EXPERIMENTAL RESULTS}

Figure~\ref{fig:RXTH} shows $Z_s$  as a function of magnetic field up to 15 T at several different temperatures. 
Here, the longitudinal axis of the lower panel in Fig.~\ref{fig:RXTH} ($\Delta R_s$) means the change from the zero-field 
value, defined as $\Delta R_s \equiv R_s(B) - R_s (0)$. 
Continuous lines and discrete open marks represent the results from swept $H$ measurements and swept $T$ measurements, 
respectively. 
Since both data sets were consistent with each other, extrinsic effects caused by an inhomogeneous field distribution in 
the data with swept $H$ measurements did not affect the essential features of the data discussed below. 
\begin{figure}[bp]
\begin{center}
\includegraphics[height=22pc]{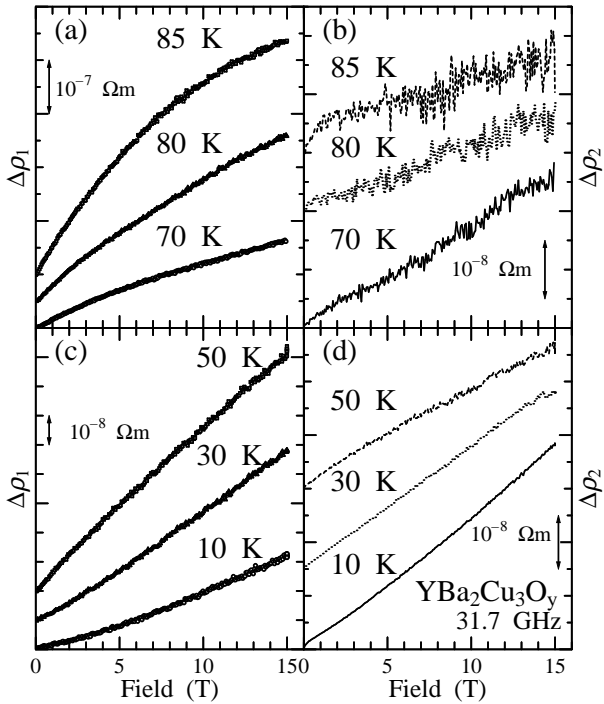}
\end{center}
\caption{Magnetic field dependence of (a) $\Delta \rho_1$ at higher temperatures, (b) $\Delta \rho_2$ at higher 
temperatures, (c) $\Delta \rho_1$ at lower temperatures, and (d) $\Delta \rho_2$ at lower temperatures, at 31.7 GHz. The 
origin of each curve has been shifted arbitrarily. }
\label{fig:RHO}
\end{figure}

In Fig.~\ref{fig:RXTH}, the magnetic field dependence of $R_s$ and $X_s$ is roughly proportional to $B^{1/2}$ over the 
whole range, which is expected in the high field region of the C-C formula. 
To see this more clearly, we plot the data in terms of the complex resistivity, $\Delta \rho_1 \equiv \rho_1 (B) - \rho_1 
(0)$, and $\Delta \rho_2 \equiv \rho_2(B)-\rho_2(0)$. 
In Fig.~\ref{fig:RHO}, we plot the magnetic field dependence of $\Delta \rho_1$ and $\Delta \rho_2$ at several fixed 
temperatures. 
If $R_s (0) \ll R_s (B)$, $\Delta \rho_1$, $\Delta \rho_2$ were expressed as $\Delta \rho_1 \simeq 2\Delta R_s X_s/\mu 
\omega$, and $\Delta \rho_2 \simeq (\Delta(X_s^2) - \Delta R_s^2)/\mu \omega$, respectively. 
According to Eqs.~(\ref{eq:crho}) and (\ref{eq:lam}), within the low temperature approximation, both $\Delta \rho_1$ and 
$\Delta \rho_2$ were proportional to a magnetic field, $B$. 
As shown in the lower two panels of Fig.~\ref{fig:RHO} ((c), and (d)), both $\Delta \rho_1$ and $\Delta \rho_2$ change 
linearly as a function of $B$ at low temperatures below 60 K. 
These behaviors suggested that the C-C model was appropriate to describe the vortex dynamics in YBCO. 
However, at higher temperatures above 70 K, as was clearly seen in $\Delta \rho_1$ (Fig.~\ref{fig:RHO} (a)), there is a 
clear deviation from the $B$-linear behavior. 
This deviation became more prominent with increasing temperature. 
In the high temperature region, thermal effects on the vortices cannot be neglected, and the low temperature approximation 
will no longer accurate. 
To obtain precise values for the vortex parameters in this regime, analysis including flux creep, and/or the $B$-dependent 
behavior of the vortex dynamics parameters is needed. 
\begin{figure}[tbp]
\begin{center}
\includegraphics[height=25pc]{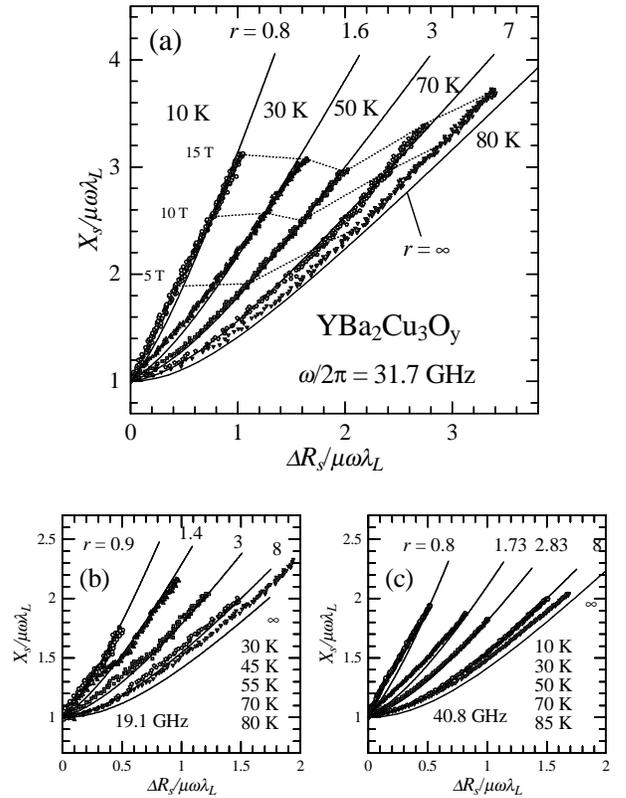}
\end{center}
\caption{(a) The impedance plane plot of the data at 31.7 GHz at various temperatures. The solid lines represent fitting 
curves to the vortex dynamics model (Eqs.~(\ref{eq:rs}) and (\ref{eq:xs})), together with the resulting $r$ values.  
Actual magnetic-field values of each line are shown as dotted lines (5, 10, and 15 T, respectively). 
Similar plots based on the data at 19.1 GHz up to 5 T, and at 40.8 GHz up to 7 T are shown in (b), and (c), respectively. 
}
\label{fig:IPDAT}
\end{figure}

In Fig.~\ref{fig:IPDAT} (a), we plot the magnetic-field dependence of $Z_s$ shown in Fig.~\ref{fig:RXTH} in the impedance 
plane. 
Results from the measurements at 19.1 GHz up to 5 T, and those at 40.8 GHz up to 7 T are also shown in 
Fig.~\ref{fig:IPDAT} (b), and (c). 
All data show a clear concave upward behavior, which suggests that we can analyze the data in terms of the vortex 
dynamics. 
The fitted curve based on Eqs.~(\ref{eq:rs}) and (\ref{eq:xs}), together with the resulting $r$ value, are also shown in 
Fig.~\ref{fig:IPDAT}. 
Indeed, the result of the fit is very good in agreement. 
Thus, we can determine $\omega_p$ and $\eta$, by the method described in the previous section. 
\begin{figure}[tbp]
\begin{center}
\includegraphics[height=25pc]{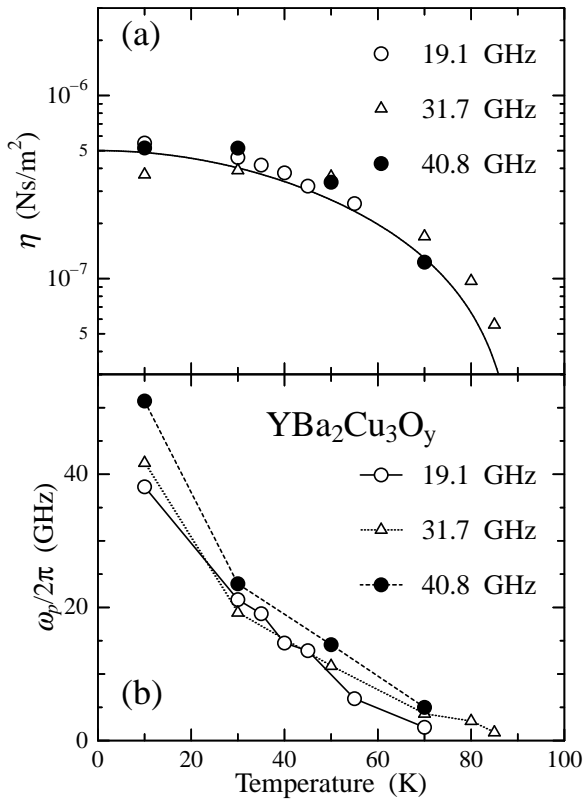}
\end{center}
\caption{Temperature dependence of the viscosity, $\eta$ (a), and the pinning frequency, $\omega_p$ (b), estimated from 
the analysis (see text). Solid curve in the panel (a) are a approximate expression, $\eta(T) = \eta (0) (1-t^2)/(1+t^2)$. 
}
\label{fig:ETA}
\end{figure}

In Fig.~\ref{fig:ETA}, we show the temperature dependence of the viscosity, $\eta$ and the pinning frequency, $\omega_p$. 
$\eta$ decreases with increasing temperature. 
The observed temperature dependence of $\eta$ was found to scale with that of the upper critical field, $B_{c2}$. 
In fact, if we assume that the temperature dependence of $B_{c2}$ as $B_{c2} = B_{c2}(0)(1-t^2)/(1+t^2)$, where $t=T/T_c$, 
and $B_{c2}(0) = 80$ T and $\rho_n \sim 3 \times 10^{-7}$~$\Omega$m, the solid curve in  Fig.~\ref{fig:ETA} (a) was 
obtained. 
The data are almost in agreement with this approximate curve. 
Deviations might be caused by the temperature dependence of the normal state resistivity, $\rho_n$, and/or an ambiguity in 
the high temperature region that will be discussed later. 
The pinning frequency $\omega_p$ decreased with increasing temperature due to thermal effects. 
Such behavior in the temperature dependence of $\eta$ and $\omega_p$ is almost consistent with the previous 
reports~\cite{PambiIEEE93,Morgan94,Revenaz94,GoloPRB94}. 
As was already mentioned, Eqs.~(\ref{eq:rs}) and (\ref{eq:xs}) are approximate expressions in the low temperature region, 
and in fact, the fit was not so good above 70 K. 
Therefore, the absolute values of these parameters at high temperature (above 70 K) might have some ambiguity. 
However, in the low temperature region, on which we mainly focus in this study, these parameters can be well defined 
within this approximation.

It was surprising that the obtained values of $\eta$ and $\omega_p$ from the measurements at different frequencies are 
almost the same. 
This fact indicates that the frequency dependence of $Z_s$ was also in good agreement with the vortex dynamics description 
with frequency-independent values for $\eta$ and $\omega_p$. 
The viscosity $\eta$ and the depinning frequency $\omega_p$ at 10 K were found to be $\sim$ 4 - 5 $\times$ 10$^{-7}$ 
Ns/m$^2$, and $\sim$ 40 - 50 GHz, respectively.
Since we kept $\mathbf{J} = {\rm const.}$ in our experiments (where $\mathbf{J}$ is the amplitude of the microwave 
current), the viscosity obtained from the data corresponds to $\eta_{\rm eff}$ in Eq.~(\ref{eq:etaeff}). 
In YBCO, the carrier concentration $n$ was proposed to be $n =$ 0.25 per CuO$_{\rm 2}$ plane~\cite{TokuraPBR}.
By using Eq.~(\ref{eq:etatau}), we estimate that $\omega_0 \tau$ is $\sim$ 0.3 - 0.5. 
Since the $\omega_p$ was found to be 40 - 50 GHz, even in the measurements in the microwave and millimeterwave regions, we 
cannot conclude that the pinning is negligible. 
In that sense, the analysis in Ref.~\cite{Matsuda94}, leading to the conclusion of the superclean core, was incorrect. 
Our results confirmed that the measurement of both the real part and the imaginary part is essential for HTSC. 
\begin{figure}[bp]
\begin{center}
\includegraphics[height=22pc]{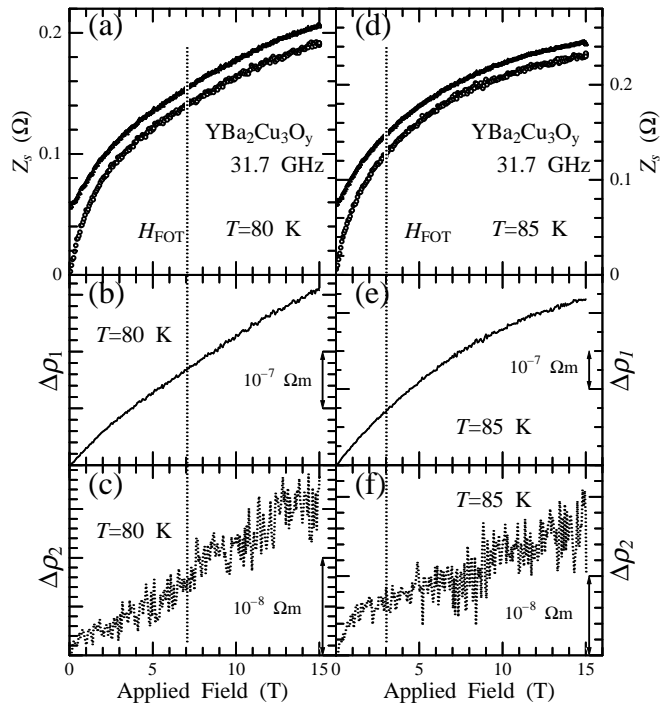}
\end{center}
\caption{Magnetic field dependence of $Z_s$, $\Delta \rho_1$, and $\Delta \rho_2$ around the FOT of the vortex lattice. 
Left three panels ((a), (b), and (c)) and right three panels ((d), (e), and (f)) are the data at 80 K, and 85 K, 
respectively. Dotted lines in each panel represent the FOT field at each temperature, which was estimated from the 
resistivity data of sample from the same batch. }
\label{fig:FOT}
\end{figure}

Next, let us move to the behavior of $Z_s$ around the FOT in the vortex lattice. 
Figure~\ref{fig:FOT} shows the magnetic-field dependence of $Z_s$, $\Delta \rho_1$, and $\Delta \rho_2$ up to 15 T at 80 
K, and 85 K. 
The dashed lines in each panel show the field corresponding to the FOT, which was determined by the resistivity data of a 
sample with the same doping level~\cite{Nishizaki98}. 
We could not find any obvious anomaly at the FOT in YBCO within our experimental resolution. 
This result is in contrast to what was observed in BSCCO with the most remarkable being the anisotropy. 
One possible reason for the absence is that the change of London penetration depth, $\lambda_L$, is too small to be 
detected. 
In YBCO, the anisotropy parameter $\gamma$ is smaller than that of BSCCO and the phase transition occurs at rather higher 
fields, which are the order of several tesla. 
In such high fields, the density of vortices is so large that the contribution of moving vortices to $Z_s$ may hide the 
subtle change of the superfluid density. 
However, if the anomaly is supposed to be as the same magnitude as in BSCCO, it should be $\Delta X_s/X_s \sim$ 4 - 5\% , 
which is well above the measurement noise level ($\le$ 1m$\Omega$, $\sim$ 1\%). 
We will discuss the reasons for this difference later. 

\section{DISCUSSION}

We found that the change in the resistivity was linear in $B$. 
We also clarified that the flux-flow in YBCO was well described in terms of the Coffey-Clem unified theory of vortex 
motion with $B$-independent, and frequency-independent vortex dynamics parameters, and essentially not different from that 
in CSC. 
As will be seen below, this is consistent with theoretical considerations for $d$-wave 
superconductors~\cite{KopVolPL97,Makhlin}

Theoretically, flux-flow was first discussed by Bardeen and Stephen.~\cite{BarSte}.
They calculated the flux flow resistivity in the dirty limit, $\delta E \sim \hbar/\tau > \Delta_0$ for $s$-wave 
superconductors.
After that, Larkin and Ovchinikov~\cite{LarOvcJL76} showed results for the case of a moderately clean core with $s$-wave 
gap symmetry, and Kopnin and Kravtsov~\cite{KopKraJL76} calculated the conductivity and the Hall angle taking the Hall 
effects into account. 
Recently, Kopnin and Volovik~\cite{KopVolPL97}, and Makhlin~\cite{Makhlin} described the flux flow in the clean, $d$-wave 
case. 
They showed that, in the moderately clean limit, the QPs outside the core did not make an important contribution to the 
flux-flow conductivity. 
The expression for the flux-flow conductivity in a $d$-wave superconductor was similar to that for an $s$-wave 
superconductor in the moderately clean regime. 
This means that the dissipative process mainly occurs at the boundary of the core, and is independent of the details of 
the electronic structure inside the core in the clean regime. 
Again, our results suggest that the flux-flow resistivity of YBCO was $B$-linear, which is not so much different from that 
in a conventional $s$-wave superconductor. 
This result is quite consistent with the theoretical description of the flux-flow in a $d$-wave superconductor. 

Next, we found that the vortex viscosity $\eta$ was 4 - 5 $\times$ 10$^{-7}$ Ns/m$^2$ at 10 K, which means that the vortex 
core of YBCO is in the moderately clean regime. 
As was described earlier, the electronic structure of vortices in HTSC is unusual because of (1) the $d$-wave symmetry of 
the energy gap, and (2) the quantum nature of the levels in the core. 
However, most of the existing theories took the quasiclassical approach to calculate the electronic structure of the core. 
In HTSC, as $k_F \xi$ is $\sim 10$ and $\hbar \omega_0$ is $\sim 10$ K, the vortex core is likely to approach the quantum 
limit ($k_F \xi \sim 1$). 
In the quantum limit, since the minimal gap $\hbar \omega_0$ is nearly equal to the energy gap $\Delta_0$, the distinction 
between QPs inside and outside the core becomes meaningless. 
Larkin and Ovchinikov~\cite{LarOvcPB98}, and Koulakov and Larkin~\cite{KouLarPB99} calculated the $I$-$V$ characteristics and microwave absorption in the quantum, clean, $s$- wave cases, respectively. 
However, to our knowledge, there was no theoretical prediction on the flux flow state in the quantum, moderately clean, 
$d$- wave case. 
All of these estimations mean that the vortex core in HTSC is in the marginal region from the different points of view 
discussed above. 

Next, we will discuss the absence of any anomaly in the microwave response in YBCO at the FOT. 
In BSCCO, the rapid increase of $X_s$ at the FOT was observed in both optimally- and over-doped 
samples~\cite{HanaguriMOS}. 
Note that no obvious anomaly was observed in $R_s$ at the FOT. 
This provides strong evidence that the origin of the anomaly could be considered as arising from the change of the 
superfluid density. 

One possibility for the absence of the anomaly is that the mechanism of the FOT is different between BSCCO and YBCO. 
A key difference between YBCO and BSCCO is anisotropy. 
According to~\cite{GlatKoshe}, the characteristic field, $B_{cr}$, which separates the regions of three-dimensional and 
two-dimensional vortex behavior, was expressed as $B_{cr} \sim \phi_0/\gamma^2 d^2$, where $\gamma^2 \equiv m_c/m_{ab}$ 
($m_{ab}$ and $m_c$ are the anisotropic effective mass in the $ab$ plane and along the $c$ axis, respectively), is the 
anisotropy parameter and $d$ is the distance between adjacent CuO$_2$ layers. 
Typical values for $\gamma^2$ are 50 and $\simeq$ 20,000 in YBCO and BSCCO, respectively. 
While the vortices in BSCCO have a 2D-like character, the vortex lattice in YBCO is rather 3D-like. 
From this point of view, our results suggest that the rapid increase of $X_s$ due to the change of superfluid density at 
FOT would be observed only in a highly anisotropic two-dimensional vortex system like BSCCO. 
In BSCCO, it has been established that the interlayer coupling changes dramatically at the FOT~\cite{ShibauchiPRL}. 
This change of interlayer coupling can cause a change in the current distribution in the sample on a macroscopic scale, 
which possibly produces the change in the in-plane properties such as $R_s$ and $X_s$~\cite{Bulaev}
However, if it is the case for the anomaly we have discussed, a similar anomaly should appear also in $R_s$, which is 
quite different from what we observed. 
Thus, we do not believe that such a change is responsible for the anomaly.

Another possibility is the difference of energy between the two materials. 
In YBCO, the complex resistivity as a function of magnetic field was measured through the FOT line up to 50 
MHz~\cite{WuOngPL97}. 
They reported definite changes in both real and imaginary parts of the complex resistivity, and interpreted the results as 
an abrupt change of the viscosity and the collapse of the shear modulus due to the melting of the vortex lattice. 
However, in the microwave response, no sign of the FOT was observed. 
We speculate that the signal due to the change of the electronic state, may exist in another frequency window, lower or 
higher than the microwave range. 

Finally, we briefly mention the attempt to detect the new resonance.
Quite recently, Balatsky {\it et al.} predicted the existence of a new collective mode~\cite{Balatsky00} whose frequency 
is tunable by magnetic field. 
According to~\cite{Balatsky00}, $\omega_{cl} \sim 72 \times H/H_{c2}$ GHz.
Assuming that $H_{c2} \sim$ 100 T, $\omega_{cl}$ at 15 T is $\sim$ 11 GHz, which is lower than our measurement 
frequencies.
If $H_{c2}$ shows a usual BCS-like $T$ dependence, the resonance originating from this mode would exist in a rather high 
$T$ region below 15 T.
However, we could not observe such as a resonance structure in our measurements of $Z_s(B)$, and $Z_s(T)$.
If the lifetime of this mode were strongly influenced by thermal smearing, this mode would be observed in the rather low 
temperature region.

In order to investigate the behavior around the FOT and find the possible collective excitation in HTSC, a measurement at 
several GHz under high magnetic field is now in progress. 

\section{CONCLUSION}

The microwave electromagnetic response in the mixed state of YBa$_2$Cu$_3$O$_y$ (YBCO) was measured in order to 
investigate the electronic state around the vortex core. 
The magnetic-field dependence of the complex surface impedance at low temperature was in good agreement with a general 
vortex dynamics description assuming that the field-independent viscous damping force and the linear restoring force acted 
on vortices. 
In other words, both the real and imaginary parts of the complex resistivity, $\rho_1$, and $\rho_2$ were linear in $B$. 
This is explained by theories for $d$-wave superconductors. 
From analysis based on the Coffey-Clem description of the complex penetration depth, we estimated that the vortex 
viscosity $\eta$ at 10 K was 4 - 5 $\times$ 10$^{-7}$ Ns/m$^2$. 
This value corresponds to $\omega_0 \tau \sim$ 0.3 - 0.5, where $\omega_0$ and $\tau$ are the minimal gap frequency and 
the quasiparticle lifetime in the vortex core, respectively. 
These results definitely show that the core of YBCO is in the moderately clean regime. 
Investigation of the moderately clean core of HTSC is significant because physically new effects are expected from the 
characteristic $d$-wave and quantum nature of cuprate superconductors. 
The behavior of $Z_s$ as a function of $B$ across the FOT was also investigated. 
Unlike in Bi$_2$Sr$_2$CaCu$_2$O$_y$ (BSCCO), a distinct anomaly was not observed around the FOT in YBCO. 
Our results suggest that the rapid increase of $X_s$ due to the change of superfluid density at the FOT would be observed 
only in a highly anisotropic two-dimensional vortex system like BSCCO. 

\begin{acknowledgments}
We are grateful to Y. Kato, A. V. Balatsky, and L. N. Bulaevskii for fruitful discussions and valuable comments, and D. G. 
Steel for a critical reading of the manuscript. 
Authors would like to thank R. Matsuo for his technical support in the construction of the equipment used in this study. 
This work is partly supported by a Grant-in-Aid for Scientific Research on Priority Area (A) No. 258, ``Vortex Electronics" sponsored by the Ministry of Education, Culture, Sports, Science, and Technology. 
Y. T. thanks the Japan Society for the Promotion of Science for financial support. 
\end{acknowledgments}


\end{document}